\documentclass[12pt]{article}
\usepackage{amsfonts}
\usepackage{amssymb}
\usepackage{bm}
\usepackage[dvips]{epsfig}
\usepackage{graphicx}
\begin{document}

\renewcommand{\theequation}{\thesection.\arabic{equation}}

\title{The deformations of Whitham systems and Lagrangian formalism.}

\author{A.Ya. Maltsev}

\date{
\centerline{L.D.Landau Institute for Theoretical Physics,}
\centerline{119334 ul. Kosygina 2, Moscow, maltsev@itp.ac.ru}}

\maketitle

\begin{abstract}
We consider the Lagrangian formalism of the deformations of
Whitham systems having Dubrovin-Zhang form. 
As an example the case of modulated one-phase
solutions of the non-linear "V-Gordon" equation is considered.
\end{abstract}

\section{Introduction.}

 This paper is a continuation of the paper \cite{deform1}
connected with the deformations of the hyperbolic Whitham systems. 
The method of deformations of Whitham
systems suggested in \cite{deform1} is connected with the slow
modulations of $m$-phase quasiperiodic solutions

\begin{equation}
\label{qpsol}
\varphi^{i}(x,t) \,\,\, = \,\,\,
\Phi^{i} \left({\bf k}({\bf U}) x \, + \,
\bm{\omega}({\bf U}) t \, + \, \bm{\theta}_{0}, \, {\bf U} 
\right) \,\,\, , \,\,\,\,\, i = 1, \dots , n
\end{equation}
of some system

\begin{equation}
\label{insyst}
F^{i} \left( \bm{\varphi}, \bm{\varphi}_{t}, \bm{\varphi}_{x},
\dots \right) \,\,\, = \,\,\, 0 
\,\,\, , \,\,\,\,\, i = 1, \dots , n
\end{equation}

 Here the functions $\Phi^{i}(\bm{\theta}, {\bf U})$ are
$2\pi$-periodic functions w.r.t. each $\theta^{\alpha}$,
$\alpha = 1, \dots , m$, the values
$\bm{\theta}_{0} = (\theta^{1}_{0}, \dots , \theta^{1}_{m})$
are arbitrary initial phase shifts and the variables
${\bf U} = (U^{1}, \dots , U^{N})$ play the role of the
parameters of $m$-phase solutions. The functions
$\bm{\omega}({\bf U}) = (\omega^{1}({\bf U}), \dots , 
\omega^{m}({\bf U}))$ and
${\bf k}({\bf U}) = (k^{1}({\bf U}), \dots ,
k^{m}({\bf U}))$ are the "frequencies" and the
"wave numbers" of the solution (\ref{qpsol}) such that the
functions $\Phi^{i}(\bm{\theta}, {\bf U})$ satisfy the system

\begin{equation}
\label{phasesyst}
F^{i} \left( \bm{\Phi}, \omega^{\alpha} ({\bf U}) 
\bm{\Phi}_{\theta^{\alpha}}, k^{\beta} ({\bf U}) 
\bm{\Phi}_{\theta^{\beta}}, \dots \right) \,\,\, = \,\,\, 0
\end{equation}
for every $\bm{\theta}$ and ${\bf U}$.

 In Whitham method the small parameter $\epsilon$ is
introduced such that $X = \epsilon x$ and $T = \epsilon t$
play the role of the "slow" coordinates in $(x,t)$-space.
The corresponding slowly modulated solutions of (\ref{qpsol})
are represented in the asymptotic form

\begin{equation}
\label{modsol}
\phi^{i}(\bm{\theta},x,t) \,\,\, = \,\,\,
\sum_{k \geq 0} \Psi_{(k)}^{i} \left(
{{\bf S}(X,T) \over \epsilon} + \bm{\theta}_{0}, X, T \right)
\,\, \epsilon^{k}
\end{equation}
where all $\bm{\Psi}_{(k)}(\bm{\theta},X,T)$ are $2\pi$-periodic
w.r.t. each $\theta^{\alpha}$ functions. The function
${\bf S}(X,T) = (S^{1}(X,T), \dots , S^{m}(X,T))$ is the
"modulated phase" of the solution (\ref{modsol}). The function
$\bm{\Psi}_{(0)}(\bm{\theta},X,T)$ satisfies the system
(\ref{phasesyst}) and belongs to the family 
$\Lambda = \{\bm{\Phi}(\bm{\theta} + \bm{\theta}_{0}, {\bf U})\}$
at every fixed $X$ and $T$. We have then

\begin{equation}
\label{psi0}
\bm{\Psi}_{(0)} (\bm{\theta},X,T) \,\,\, = \,\,\,
\bm{\Phi} \left( \bm{\theta} + \bm{\theta}_{0}(X,T), {\bf U}(X,T)
\right)
\end{equation}
and

$$S^{\alpha}_{T} \, = \, \omega^{\alpha}({\bf U}(X,T)) \,\,\, ,
\,\,\,\,\, S^{\alpha}_{X} \, = \, k^{\alpha}({\bf U}(X,T)) $$
as follows from the substitution of (\ref{modsol}) in the system
(\ref{insyst}).

 The functions $\bm{\Psi}_{(k)} (\bm{\theta},X,T)$ are defined
from the linear systems

\begin{equation}
\label{ksyst}
{\hat L}^{i}_{j[{\bf U}, \bm{\theta}_{0}]}(X,T) \,\,
\Psi_{(k)}^{j} (\bm{\theta},X,T) \,\,\, = \,\,\, 
f_{(k)}^{i} (\bm{\theta},X,T)
\end{equation}
where ${\hat L}^{i}_{j[{\bf U}, \bm{\theta}_{0}]}(X,T)$ 
is a linear operator given by the linearizing of the system
(\ref{phasesyst}) on the solution (\ref{psi0}). The resolvability
conditions of the system (\ref{ksyst}) can be written as the
orthogonality conditions of the functions 
${\bf f}_{(k)} (\bm{\theta},X,T)$ to all the "left eigen vectors"
(the eigen vectors of adjoint operator) 
$\bm{\kappa}^{(q)}_{[{\bf U}(X,T)]} 
(\bm{\theta} + \bm{\theta}_{0}(X,T))$ of the operator 
${\hat L}^{i}_{j[{\bf U}, \bm{\theta}_{0}]}(X,T)$ corresponding to
zero eigen-values. The resolvability conditions of (\ref{ksyst})
for $k = 1$ together with

$$k^{\alpha}_{T} = \omega^{\alpha}_{X} $$
give the Whitham system for $m$-phase solutions of (\ref{insyst})
playing the central role in the slow modulations approach.

 Like in \cite{deform1} we will assume here that the parameters
$({\bf k}, \bm{\omega})$ can be considered as the independent
parameters on the family $\Lambda$ such that the full set of
independent parameters ${\bf U}$ (except initial phases
$\theta^{\alpha}_{0}$) can be represented in the form
$({\bf k}, \bm{\omega}, {\bf n})$ where ${\bf k}$ and
$\bm{\omega}$ are the wave numbers and the frequencies of the 
solution and ${\bf n} = (n^{1}, \dots n^{s})$ are some additional
parameters (if they exist).

 Easy to see that the functions

$$\bm{\Phi}_{\theta^{\alpha}} (\bm{\theta} + \bm{\theta}_{0},
{\bf k}, \bm{\omega}, {\bf n}) \,\,\, , \,\,\,\,\,
\alpha = 1 , \dots , m $$
and

$$\bm{\Phi}_{n^{l}} (\bm{\theta} + \bm{\theta}_{0}, 
{\bf k}, \bm{\omega}, {\bf n}) \,\,\, , \,\,\,\,\,
l = 1 , \dots , s $$
give the  eigen-vectors of the operator 
${\hat L}^{i}_{j[\bm{\theta}_{0},{\bf k},\bm{\omega},{\bf n}]}$
corresponding to zero eigen-values.

 Let us give also the definition of the full regular family
of $m$-phase solutions of (\ref{insyst}).\footnote{This 
definition corresponds to the "weak" definition of full
regular family of $m$-phase solutions given in \cite{deform1}.}

\vspace{0.5cm}

{\bf Definition 1.1.}

{\it
We call the family $\Lambda$ the full regular family of 
$m$-phase solutions of (\ref{insyst}) if

1) The functions 
$\bm{\Phi}_{\theta^{\alpha}} (\bm{\theta}, 
{\bf k}, \bm{\omega}, {\bf n})$, 
$\bm{\Phi}_{n^{l}} (\bm{\theta}, {\bf k}, \bm{\omega}, {\bf n})$
are linearly independent and give (for generic ${\bf k}$ and
$\bm{\omega}$) the full basis in the kernel of the operator
${\hat L}^{i}_{j[\bm{\theta}_{0},{\bf k},\bm{\omega},{\bf n}]}$;

2) The operator 
${\hat L}^{i}_{j[\bm{\theta}_{0},{\bf k},\bm{\omega},{\bf n}]}$
has (for generic ${\bf k}$ and $\bm{\omega}$) exactly $m + s$ 
linearly independent "left eigen vectors"

$$\bm{\kappa}^{(q)}_{[{\bf U}]} (\bm{\theta} + \bm{\theta}_{0})
\,\,\, = \,\,\, 
\bm{\kappa}^{(q)}_{[{\bf k}, \bm{\omega}, {\bf n}]} 
(\bm{\theta} + \bm{\theta}_{0}) $$
depending on the parameters ${\bf U}$ in a smooth way and 
corresponding to zero eigen-values.
}

\vspace{0.5cm}

 It can be shown that for full regular family $\Lambda$ the
corresponding Whitham system puts the restrictions only on the
functions

$${\bf U}(X,T) \,\,\, = \,\,\, \left(
{\bf k} (X,T), \bm{\omega} (X,T), {\bf n} (X,T) \right) $$
having the form

$$k^{\alpha}_{T} \,\,\, = \,\,\, \omega^{\alpha}_{X} $$
\begin{equation}
\label{whithsyst}
C^{(q)}_{\nu} ({\bf U}) \, U^{\nu}_{T} \,\, - \,\,
D^{(q)}_{\nu} ({\bf U}) \, U^{\nu}_{X} \,\, = \,\, 0
\end{equation}
($q = 1, \dots , m+s$, $\nu = 1, \dots , N = 2m+s$) and does 
not include the initial phase shifts $\theta^{\alpha}_{0}(X,T)$.

\vspace{0.5cm}

{\bf Definition 1.2.}

{\it
Let us call the Whitham system (\ref{whithsyst}) non-degenerate
hyperbolic Whitham system if:

1) The system (\ref{whithsyst}) is resolvable with respect to the 
time derivatives of parameters $U^{\nu}$ and can be written in the
form

\begin{equation}
\label{HTsyst}
U^{\nu}_{T} \,\,\, = \,\,\, V^{\nu}_{\mu} ({\bf U}) \,
U^{\mu}_{X} \,\,\,\,\, , \,\,\,\,\,\,\,\,
\nu , \mu = 1, \dots , N
\end{equation}

2) The matrix $V^{\nu}_{\mu} ({\bf U})$ has 
$N$ linearly independent real eigen-vectors with real 
eigen-values.
}

\vspace{0.5cm}

 Provided that the system (\ref{whithsyst}) is satisfied we can find 
the first correction $\bm{\Psi}_{(1)} (\bm{\theta}, X, T)$ 
in the solution
(\ref{modsol}) modulo the linear combination of the functions
$\bm{\Phi}_{\theta^{\alpha}} (\bm{\theta} + \bm{\theta}_{0},
{\bf k}, \bm{\omega}, {\bf n})$,
$\bm{\Phi}_{n^{l}} (\bm{\theta} + \bm{\theta}_{0},
{\bf k}, \bm{\omega}, {\bf n})$. In general scheme we try to
find recursively the higher order corrections
$\bm{\Psi}_{(k)} (\bm{\theta}, X, T)$ from the linear systems 
(\ref{ksyst}). The functions $\theta^{\alpha}_{0} (X, T)$ and
the freedom in the determination of the functions
$\bm{\Psi}_{(k)} (\bm{\theta}, X, T)$ are used to satisfy the 
compatibility conditions of the systems (\ref{ksyst}) in higher
orders of $\epsilon$, so we get the recursive restrictions 
on the corresponding parameters.\footnote{For systems
(\ref{insyst}) having non-degenerate hyperbolic Whitham systems 
it was shown in \cite{deform1} that the asymptotic series
(\ref{modsol}) can be constructed globally in $X$ up to the
moment of the breakdown of corresponding solution of Whitham 
system.} The solution of the Whitham system (\ref{whithsyst})
(or (\ref{HTsyst})) is considered usually as the central
point of the procedure which defines the global properties of
the modulated solution. Let us also mention the well known
fact that the Whitham systems corresponding to the integrable
systems (\ref{insyst}) possess also the integrability properties.

The first consideration of dispersive corrections to Whitham
systems were made in \cite{AblBenny} 
(see also \cite{Abl1}-\cite{Abl2}) where the multi-phase 
Whitham was also first discussed. As was pointed out in
\cite{AblBenny} the dispersive corrections can naturally arise
in the Whitham method both in one-phase and multi-phase situations.

 Here we consider the deformations of Whitham systems
(\ref{HTsyst}) having the form of Dubrovin-Zhang deformations
of Frobenius manifolds (\cite{DubrZhang1,DubrZhang2}). 
The problem is thus connected with B.A. Dubrovin problem of 
deformations of Frobenius manifolds corresponding to Whitham
systems of integrable hierarchies.
 
 According to Dubrovin-Zhang approach we call the deformation
of the Whitham system (\ref{whithsyst}) (or (\ref{HTsyst}))
the expression containing the system (\ref{whithsyst}) 
(or (\ref{HTsyst})) as the leading term and the infinite
number of "dispersive" corrections containing the higher
($T$ and $X$) derivatives of the parameters $U^{\nu}$
and polynomial with respect to all derivatives of $U^{\nu}$.
For "non-degenerate hyperbolic" Whitham systems
(\ref{HTsyst}) it is natural to express recursively all the
higher $T$-derivatives of parameters $U^{\nu}$ in terms of their
$X$-derivatives and represent the deformation of the Whitham
system in the evolution (Dubrovin-Zhang) form

\begin{equation}
\label{defsyst}
U^{\nu}_{T} \,\,\, = \,\,\, V^{\nu}_{\mu} ({\bf U}) \,
U^{\mu}_{X} \,\, + \,\, \sum_{k \geq 2}
\zeta^{\nu}_{(k)} ({\bf U}, {\bf U}_{X}, {\bf U}_{XX}, \dots )
\end{equation}
 
 We require now that all $\zeta^{\nu}_{(k)}$ satisfy the following 
conditions

\vspace{0.5cm}

1) All $\zeta^{\nu}_{(k)}$ are polynomial in derivatives
${\bf U}_{X}, {\bf U}_{XX}, \dots $

\vspace{0.5cm}

2) The term $\zeta^{\nu}_{(k)}$ has degree $k$ according to the 
following gradation:

All the functions $f({\bf U})$ have degree $0$;

The derivatives $U^{\nu}_{kX}$ have degree $k$;

The degree of the product of functions having certain degrees
is equal to the sum of their degrees.

\vspace{0.5cm}

 Let us say that the deformation of the Whitham system having
the form (\ref{defsyst}) with all the conditions formulated
above is represented in Dubrovin - Zhang form.

\vspace{0.5cm}

 Let us describe briefly the main features of deformation
procedure of Whitham systems suggested in \cite{deform1}.
We will write here the parameters ${\bf U}$ in the form
$({\bf k}, \bm{\omega}, {\bf n})$. According to \cite{deform1}
we have to make first the "right renormalization" of the
functions $S^{\alpha}(X,T)$ and $n^{l}(X,T)$ arising in
Whitham method. Namely, we allow the regular 
$\epsilon$-dependence of the functions $S^{\alpha}(X,T)$,
$n^{l}(X,T)$ having the form of the infinite series

$$S^{\alpha} (X,T,\epsilon) \,\,\, = \,\,\,
\sum_{k \geq 0} \epsilon^{k} \,\, S_{(k)}^{\alpha} (X,T) $$

$$n^{l} (X,T,\epsilon) \,\,\, = \,\,\,
\sum_{k \geq 0} \epsilon^{k} \,\, n_{(k)}^{l} (X,T) $$

 However, we require now that the function

\begin{equation}
\label{psiappr}
\bm{\phi}_{(0)} ( \bm{\theta} , X, T, \epsilon )
\,\,\, = \,\,\, \bm{\Phi} \left(
{{\bf S}(X,T,\epsilon) \over \epsilon} \, + \, \bm{\theta}, \,
{\bf S}_{X} (X,T,\epsilon), \, {\bf S}_{T} (X,T,\epsilon), \,
{\bf n} (X,T,\epsilon) \right)
\end{equation}
gives the "best possible" approximation to the full asymptotic
solution (\ref{modsol}). More precisely, we require that the
function (\ref{psiappr}) satisfies the following conditions

\begin{equation}
\label{norm1}
\int_{0}^{2\pi} \!\!\! \dots \int_{0}^{2\pi} \sum_{i=1}^{n}
\phi^{i}_{(0)\theta^{\alpha}} (\bm{\theta},X,T,\epsilon ) \,\,
\phi^{i} (\bm{\theta},X,T,\epsilon ) \,\,
{d^{m} \theta \over (2\pi)^{m}} \,\,\, \equiv
\end{equation}
$$\equiv \,\,\,
\int_{0}^{2\pi} \!\!\! \dots \int_{0}^{2\pi} \sum_{i=1}^{n}
\phi^{i}_{(0)\theta^{\alpha}} (\bm{\theta},X,T,\epsilon ) \,\,
\phi^{i}_{(0)} (\bm{\theta},X,T,\epsilon ) \,\,
{d^{m} \theta \over (2\pi)^{m}} \,\,\, \equiv \,\,\, 0 
\,\,\,\,\, , \,\,\,\,\,\,\,\, \alpha = 1, \dots , m $$

\begin{equation}
\label{norm2}
\int_{0}^{2\pi} \!\!\! \dots \int_{0}^{2\pi} \sum_{i=1}^{n}
\phi^{i}_{(0)n^{l}} (\bm{\theta},X,T,\epsilon ) \,\,
\phi^{i} (\bm{\theta},X,T,\epsilon ) \,\,
{d^{m} \theta \over (2\pi)^{m}} \,\,\, \equiv
\end{equation}
$$\equiv \,\,\,
\int_{0}^{2\pi} \!\!\! \dots \int_{0}^{2\pi} \sum_{i=1}^{n}
\phi^{i}_{(0)n^{l}} (\bm{\theta},X,T,\epsilon ) \,\,
\phi^{i}_{(0)} (\bm{\theta},X,T,\epsilon ) \,\,
{d^{m} \theta \over (2\pi)^{m}} \,\,\,\,\, , \,\,\,\,\,\,\,\, 
l = 1, \dots , s $$
where $\bm{\phi} (\bm{\theta},X,T,\epsilon )$ is the asymptotic
solution given by (\ref{modsol}).

 After that we try to make a "re-expansion" of the asymptotic
series (\ref{modsol}) using the higher derivatives of the
"renormalized" functions ${\bf S}(X,T,\epsilon)$,
${\bf n}(X,T,\epsilon)$ instead of the parameter $\epsilon$
in the expansion. As was shown in \cite{deform1} in the case
of non-degenerate hyperbolic Whitham system (\ref{whithsyst})
it is possible to use only the $X$-derivatives of parameters
$({\bf k}, \bm{\omega}, {\bf n})$ in this expansion and put
the conditions of the form (\ref{defsyst}) on the 
"renormalized" functions ${\bf k}(X,T,\epsilon)$, 
$\bm{\omega}(X,T,\epsilon)$, ${\bf n}(X,T,\epsilon)$.
The asymptotic solution (\ref{modsol}) is represented then
as the new asymptotic expansion with respect to higher
$X$-derivatives of parameters 
${\bf U} = ({\bf k}, \bm{\omega}, {\bf n})$ according 
to the gradation introduced above.

 After the "re-expansion" we can forget in fact about the
$\epsilon$-dependence of the functions
${\bf S}(X,T,\epsilon)$, ${\bf n}(X,T,\epsilon)$ and consider
the "concrete formal solutions" of the system (\ref{insyst})
without the additional one-parametric $\epsilon$-families.
The $X$-derivatives of the slow functions
${\bf k}(X,T)$, $\bm{\omega}(X,T)$, ${\bf n}(X,T)$
play now the role of the small parameters in the expansion with
the gradation introduced above. We keep now the notations
$X$ and $T$ for spatial and time variables just to emphasize
that we consider the slow functions of $x$ and $t$. 

 The formal solution of (\ref{insyst}) will be written now in 
the form

\begin{equation}
\label{gradexp}
\phi^{i}(\bm{\theta}, X, T) \,\,\, = \,\,\,
\Phi^{i} \left( {\bf S}(X,T) + \bm{\theta}, {\bf S}_{X},
{\bf S}_{T}, {\bf n} \right) \,\, + \,\,
\sum_{k \geq 1} \Phi^{i}_{(k)}
\left( {\bf S}(X,T) + \bm{\theta}, X, T \right)
\end{equation}
where all $\Phi^{i}_{(k)} (\bm{\theta}, X, T)$ are local
expressions depending on
$({\bf k}, \bm{\omega}, {\bf n}, {\bf k}_{X}, \bm{\omega}_{X}, 
{\bf n}_{X}, \dots )$ polynomial in derivatives
$({\bf k}_{X}, \bm{\omega}_{X}, {\bf n}_{X}, \dots )$
and having degree $k$. 

 All $\bm{\Phi}_{(k)} (\bm{\theta}, X, T)$, $ k \geq 1$
satisfy the normalization conditions

\begin{equation}
\label{newnorm1}
\int_{0}^{2\pi} \!\!\! \dots \int_{0}^{2\pi} \sum_{i=1}^{n}
\Phi^{i}_{\theta^{\alpha}} (\bm{\theta}, {\bf S}_{X},
{\bf S}_{T}, {\bf n} ) \,\, \Phi^{i}_{(k)} (\bm{\theta}, X, T)
\,\, {d^{m} \theta \over (2\pi)^{m}} \,\,\, \equiv \,\,\, 0
\,\,\,\,\, , \,\,\,\,\,\,\,\, \alpha = 1, \dots , m
\end{equation}

\begin{equation}
\label{newnorm2}
\int_{0}^{2\pi} \!\!\! \dots \int_{0}^{2\pi} \sum_{i=1}^{n}
\Phi^{i}_{n^{l}} (\bm{\theta}, {\bf S}_{X},
{\bf S}_{T}, {\bf n} ) \,\, \Phi^{i}_{(k)} (\bm{\theta}, X, T)
\,\, {d^{m} \theta \over (2\pi)^{m}} \,\,\, \equiv \,\,\, 0
\,\,\,\,\, , \,\,\,\,\,\,\,\, l = 1, \dots , s
\end{equation}
according to normalization (\ref{norm1})-(\ref{norm2}).

 The functions $\bm{\Phi}_{(k)} (\bm{\theta}, X, T)$
satisfy the linear systems

\begin{equation}
\label{newksyst}
{\hat L}^{i}_{j[{\bf S}_{X},{\bf S}_{T},{\bf n}]} (X,T) \,\,
\bm{\Phi}_{(k)} (\bm{\theta}, X, T) \,\,\, = \,\,\,
{\tilde f}^{i}_{(k)} (\bm{\theta}, X, T) 
\end{equation}
analogous to (\ref{ksyst}). The systems (\ref{newksyst})
represent now all the terms having gradation $k$ after the
substitution of (\ref{gradexp}) in (\ref{insyst}).

 Let us say that the parameters $\theta^{\alpha}_{0}(X,T)$
and other additional parameters arising in Whitham method
do not appear in this approach being completely "absorbed"
by the "renormalized" functions ${\bf S} (X,T)$,
${\bf n}(X,T)$.

 The functions $\zeta^{\nu}_{(k)} ({\bf U}, {\bf U}_{X}, \dots)$
arising in (\ref{defsyst}) are defined now from the compatibility
conditions of the systems (\ref{newksyst}) in the $k$-th order.
The functions $\bm{\Phi}_{(k)} (\bm{\theta}, X, T)$ are uniquely
defined then from (\ref{newksyst}) view the conditions
(\ref{newnorm1})-(\ref{newnorm2}). The first term of the system 
(\ref{defsyst}) coincides with the right-hand part of the
corresponding Whitham system (\ref{HTsyst}).

 The full system (\ref{defsyst}) in parameters 
$({\bf k}, \bm{\omega}, {\bf n})$ can be written in the form

$$k^{\alpha}_{T} \,\,\, = \,\,\, \omega^{\alpha}_{X}$$
\begin{equation}
\label{konsyst}
\omega^{\alpha}_{T} \,\,\, = \,\,\, \sum_{k \geq 1}
\sigma^{\alpha}_{(k)} \left( {\bf k}, \bm{\omega}, {\bf n},
{\bf k}_{X}, \bm{\omega}_{X}, {\bf n}_{X}, \dots \right)
\end{equation}
$$n^{l}_{T} \,\,\, = \,\,\, \sum_{k \geq 1}
\eta^{l}_{(k)} \left( {\bf k}, \bm{\omega}, {\bf n},
{\bf k}_{X}, \bm{\omega}_{X}, {\bf n}_{X}, \dots \right) $$
where all $\sigma^{\alpha}_{(k)}$, $\eta^{l}_{(k)}$ are local
expressions in 
$({\bf k}, \bm{\omega}, {\bf n}, {\bf k}_{X}, \bm{\omega}_{X}, 
{\bf n}_{X}, \dots )$ polynomial in derivatives and having
degree $k$.

 It was shown in \cite{deform1} that the expansions 
(\ref{gradexp}) represent all the "particular" formal solutions
(\ref{modsol}) in the case of non-degenerate hyperbolic Whitham 
system (\ref{HTsyst}).

 We can consider now the family of 
slowly-modulated formal solutions of (\ref{insyst})
parameterized by the functions 
${\bf k}(X,T)$, $\bm{\omega}(X,T)$, ${\bf n}(X,T)$
(and the general initial phase $\bm{\theta}_{0}$)
satisfying the system (\ref{konsyst}).

 At the end of this Chapter let us make one more remark about 
the representation of slowly-modulated solutions of 
(\ref{insyst}) in the form (\ref{gradexp}). Namely, we can see 
that the representation (\ref{gradexp}) depends on the choice of 
the functions $\Phi^{i}(\bm{\theta},{\bf k},\bm{\omega},{\bf n})$
having "zero initial phase shifts" at every 
$({\bf k}, \bm{\omega}, {\bf n})$. In particular, the natural
change

\begin{equation}
\label{Phiprime}
\Phi^{\prime i}(\bm{\theta},{\bf k},\bm{\omega},{\bf n})
\,\,\, = \,\,\,
\Phi^{i}(\bm{\theta} + \delta ({\bf k},\bm{\omega},{\bf n}),
{\bf k},\bm{\omega},{\bf n})
\end{equation}
of the functions $\Phi^{i}$ is possible in our situation. 
The change (\ref{Phiprime}) of the function $\Phi^{i}$ gives
then another representation of the same family of formal
solutions parameterized by other functions
${\bf k}^{\prime}(X,T)$, $\bm{\omega}^{\prime}(X,T)$, 
${\bf n}^{\prime}(X,T)$. In general, the form of the system
(\ref{konsyst}) will also depend on the choice of the functions
$\Phi^{i}(\bm{\theta},{\bf k},\bm{\omega},{\bf n})$.

 For Dubrovin - Zhang approach to the classification of
integrable hierarchies the following statement plays important
role (\cite{deform1}):

 The systems (\ref{konsyst}) written for two sets of functions
$\bm{\Phi}$ and $\bm{\Phi}^{\prime}$ connected by the transformation
(\ref{Phiprime}) are connected by the "trivial transformation"
(\cite{DubrZhang1,DubrZhang2}) i.e.

 There exists a change of coordinates

$$k^{\prime \alpha} \,\,\, = \,\,\, k^{\alpha} \,\, + \,\,
\sum_{k \geq 1} K^{\alpha}_{(k)} \left( {\bf k}, \bm{\omega}, 
{\bf n}, {\bf k}_{X}, \bm{\omega}_{X}, {\bf n}_{X}, \dots \right)$$

$$\omega^{\prime \alpha} \,\,\, = \,\,\, \omega^{\alpha} \,\, + \,\,  
\sum_{k \geq 1} \Omega^{\alpha}_{(k)} \left( {\bf k}, \bm{\omega},
{\bf n}, {\bf k}_{X}, \bm{\omega}_{X}, {\bf n}_{X}, \dots \right)$$ 

$$n^{\prime l} \,\,\, = \,\,\, n^{l} \,\, + \,\,  
\sum_{k \geq 1} N^{l}_{(k)} \left( {\bf k}, \bm{\omega},
{\bf n}, {\bf k}_{X}, \bm{\omega}_{X}, {\bf n}_{X}, \dots \right)$$ 
where all $K^{\alpha}_{(k)}$, $\Omega^{\alpha}_{(k)}$, $N^{l}_{(k)}$
are polynomial in derivatives 
$({\bf k}_{X}, \bm{\omega}_{X}, {\bf n}_{X}, \dots )$
(and having degree $k$) which transforms the corresponding
systems (\ref{konsyst}) one into another.

 In this paper we will investigate the Lagrangian properties
of the systems (\ref{defsyst}) in the case when the initial
system (\ref{insyst}) can be written in "local" Lagrangian form

\begin{equation}
\label{Lagrform}
\delta \,\, \int\int {\cal L} (\bm{\varphi}, \bm{\varphi}_{t},
\bm{\varphi}_{x}, \dots ) \,\, dx \, dt \,\,\, = \,\,\, 0
\end{equation}

 We do not assume here the integrability of the system
(\ref{insyst}) so the corresponding statements will be valid
for both integrable and non-integrable cases. We will show
here that the existence of Lagrangian formalism (\ref{Lagrform})
gives in fact the convenient procedure of construction of the
deformed system (\ref{defsyst}). As the example we will 
consider the case of one-phase modulated solutions of
"V-Gordon" equation

$$\varphi_{tt} \,\, - \,\, \varphi_{xx} \,\, + \,\,
V^{\prime} (\varphi) \,\,\, = \,\,\, 0 $$
and describe the Lagrangian formalism for the corresponding
deformation of it's Whitham system up to the first
nontrivial correction.

\section{Lagrangian formalism.}
\setcounter{equation}{0}

 We will assume now that the initial system (\ref{insyst})
can be written in the Lagrangian form (\ref{Lagrform})
where the Lagrangian density 
${\cal L} (\bm{\varphi}, \bm{\varphi}_{t}, \bm{\varphi}_{x}, 
\bm{\varphi}_{tt}, \bm{\varphi}_{xx},\dots )$ is the local 
function of all variables. We will consider here the Lagrangian 
form of the system (\ref{defsyst}). Let us say that the Lagrangian
formalism for the Whitham system (\ref{whithsyst}) was
constructed by G. Whitham (\cite{whith3}) who suggested the
"averaged" Lagrangian formalism

\begin{equation}
\label{whithlag}
{\delta \over \delta {\bf S}(X,T)} \int\int {\bar {\cal L}}
\left( {\bf S}_{X^{\prime}}, {\bf S}_{T^{\prime}} \right)
\,\, d X^{\prime} \, dT^{\prime} \,\,\, = \,\,\, 0
\end{equation}
for (\ref{whithsyst}). It was also pointed out by G. Whitham
that the parameters ${\bf n}$ on the family $\Lambda$ can be
described with the aid of specific "pseudo-phases" in the
Lagrangian approach such that the system (\ref{whithsyst})
can be written in the form (\ref{whithlag}) in the general case.
The Lagrangian formalism (\ref{whithlag}) gives then the equations
(\ref{whithsyst}) written in the terms of phases and 
"pseudo-phases" as well as the conservation laws for the system 
(\ref{whithsyst}).

 Let us omit also here the parameters $(n^{1}, \dots, n^{s})$
and assume that the $m$-phase solutions of (\ref{insyst}) are
parameterized by the values $(k^{1}, \dots, k^{m})$,
$(\omega^{1}, \dots, \omega^{m})$ and the
initial phase shifts $(\theta_{0}^{1}, \dots, \theta_{0}^{m})$.
The corresponding considerations can then be easily
generalized to the more general case by the introduction
of "pseudo-phases" in Whitham's way.

 We will assume then that the system (\ref{insyst}) has a full
regular family $\Lambda$ of $m$-phase solutions with parameters
$k^{\alpha} = S^{\alpha}_{X}$, $\omega^{\alpha} = S^{\alpha}_{T}$, 
$\theta_{0}^{\alpha}$, $\alpha = 1, \dots, m$. Besides that, we
will assume that the corresponding Whitham system (\ref{whithsyst})
is non-degenerate hyperbolic system which makes the form 
(\ref{defsyst}) of the deformation of Whitham system the most
natural. The functions 
$\bm{\Phi}_{\theta^{\alpha}}(\bm{\theta},{\bf k},\bm{\omega})$
give now the basis (for generic ${\bf k}$ and $\bm{\omega}$) 
in the kernel of the operator 
${\hat L}^{i}_{j[\bm{\theta}_{0},{\bf k},\bm{\omega}]}$
introduced above. We require also that 
${\hat L}^{i}_{j[\bm{\theta}_{0},{\bf k},\bm{\omega}]}$
has exactly $m$ (for generic ${\bf k}$ and $\bm{\omega}$)
linearly independent "left eigen vectors" 
$\bm{\kappa}^{(q)}_{[{\bf k},\bm{\omega}]}(\bm{\theta} + 
\bm{\theta}_{0})$, $q = 1, \dots, m$ corresponding to zero
eigen-values. The Whitham system (\ref{whithsyst}) can be written
then in the form

\begin{equation}
\label{whssyst}
A^{(q)}_{\alpha}({\bf S}_{X}, {\bf S}_{T}) \,
S^{\alpha}_{TT} \,\, + \,\, 
B^{(q)}_{\alpha}({\bf S}_{X}, {\bf S}_{T}) \,
S^{\alpha}_{XT} \,\, + \,\,
C^{(q)}_{\alpha}({\bf S}_{X}, {\bf S}_{T}) \,
S^{\alpha}_{XX} \,\,\, = \,\,\, 0
\end{equation}
$\alpha = 1, \dots, m$, $q = 1, \dots, m$, or, equivalently
 
$$k^{\alpha}_{T} \,\,\, = \,\,\, \omega^{\alpha}_{X}$$
\begin{equation}
\label{whoksyst}
A^{(q)}_{\alpha}({\bf k}, \bm{\omega}) \, \omega^{\alpha}_{T}
\,\,\, = \,\,\, - B^{(q)}_{\alpha}({\bf k}, \bm{\omega}) \, 
\omega^{\alpha}_{X} \,\, - \,\,
C^{(q)}_{\alpha} ({\bf k}, \bm{\omega}) \, k^{\alpha}_{X}
\end{equation}
in the variables $({\bf k}, \bm{\omega})$.

 We want to get the deformation of the system (\ref{whoksyst})
in the form

$$k^{\alpha}_{T} \,\,\, = \,\,\, \omega^{\alpha}_{X}$$
\begin{equation}
\label{defsyst1}
\omega^{\alpha}_{T} \,\,\, = \,\,\, \sum_{k \geq 1}
\sigma^{\alpha}_{(k)} \left( {\bf k}, \bm{\omega}, 
{\bf k}_{X}, \bm{\omega}_{X}, \dots \right)
\end{equation}
where

$$\sigma^{\alpha}_{(1)} \,\, = \,\, - \, 
||A^{-1}||^{\alpha}_{\beta} \, C^{\beta}_{\gamma} \, 
k^{\gamma}_{X} \,\, - \,\, ||A^{-1}||^{\alpha}_{\beta} \,
B^{\beta}_{\gamma} \, \omega^{\gamma}_{X} $$
and all $\sigma^{\alpha}_{(k)}$ are polynomial in $X$-derivatives
of ${\bf k}$ and $\bm{\omega}$ and have degree $k$ according to
the gradation introduced above.

 We are trying to find a solution of (\ref{insyst}) in the form

\begin{equation}
\label{gradexp1}
\phi^{i}(\bm{\theta}, X, T) \,\, = \,\,
\Phi^{i} \left( {\bf S}(X,T) + \bm{\theta}, {\bf S}_{X}, 
{\bf S}_{T} \right) \,\, + \,\, \sum_{k \geq 1}
\Phi_{(k)}^{i} \left( {\bf S}(X,T) + \bm{\theta}, X, T \right) 
\end{equation}
where all $\phi_{(k)}^{i}(\bm{\theta}, X, T)$ are $2\pi$-periodic
w.r.t. each $\theta^{\alpha}$ functions which are local
functionals of ${\bf S}_{X}$, ${\bf S}_{T}$, ${\bf S}_{XX}$,
${\bf S}_{XT}$, ${\bf S}_{XXX}$, ${\bf S}_{XXT}$, $\dots$,
polynomial in ${\bf S}_{XX}$, ${\bf S}_{XT}$, ${\bf S}_{XXX}$, 
${\bf S}_{XXT}$, $\dots$, and having degree $k$. 

 All $\bm{\Phi}_{(k)}$ satisfy the normalization conditions

\begin{equation}
\label{norm}
\int_{0}^{2\pi} \!\!\! \dots \int_{0}^{2\pi} \sum_{i=1}^{n}
\Phi_{(k)}^{i}(\bm{\theta}, X, T) \,\,
\Phi^{i}_{\theta^{\alpha}} (\bm{\theta}, {\bf S}_{X}, 
{\bf S}_{T}) \,\, {d^{m} \theta \over (2\pi)^{m}} \,\,\,
\equiv \,\,\, 0
\end{equation}
$\alpha = 1, \dots, m$, $k \geq 1$.

 The system (\ref{defsyst1}) provides the existence of all terms
$\bm{\Phi}_{(k)}(\bm{\theta}, X, T)$ of solution (\ref{gradexp1})
satisfying (\ref{norm}) (\cite{deform1}) which are represented 
as the local functionals of 
$({\bf k}, \bm{\omega}, {\bf k}_{X}, \bm{\omega}_{X}, \dots)$ 
and are defined 
by the functions $\{k^{\alpha}(X), \omega^{\alpha}(X)\}$
at every $T$.

 We can say then that the functional representation 
(\ref{gradexp1}) gives a mapping $\tau$

$$\tau \, : \,\,\,\,\, \{{\bf S}(X), {\bf S}_{T}(X)\}
\,\, \rightarrow \,\, \{\bm{\varphi} (\bm{\theta}, X)\} $$
from the "loop space" $\{{\bf S}(X), {\bf S}_{T}(X)\}$ to the
functional space $\{\bm{\varphi} (\bm{\theta}, X)\}$. In this
approach we can define a "sub-manifold"  $Im \, \tau$ in the
functional space $\{\bm{\varphi} (\bm{\theta}, X)\}$ 
parameterized by the functional parameters 
$\{{\bf S}(X), {\bf S}_{T}(X)\}$.

 The system (\ref{defsyst1}) generates a dynamical flow on the
"sub-manifold"  $Im \, \tau$ corresponding to the system 
(\ref{insyst}). In this approach the system (\ref{defsyst1})
becomes the condition that the formal series 
(\ref{gradexp1}) (with known functionals 
$\Phi^{i}_{(k)} (\bm{\theta}, {\bf k}, \bm{\omega}, 
{\bf k}_{X}, \bm{\omega}_{X}, \dots)$) represents a solution
of (\ref{insyst}).

 We can suggest now the Lagrangian form for the system
(\ref{defsyst1}) using the Lagrangian formalism (\ref{Lagrform})
for the system (\ref{insyst}). We know that the function
$\bm{\phi} (\bm{\theta}, X, T)$ defined by (\ref{gradexp1})
should satisfy the Euler - Lagrange equation

\begin{equation}
\label{Lf1}
{\delta \over \delta \, \varphi^{i} (\bm{\theta}, X, T)}
\int_{-\infty}^{+\infty}\int_{-\infty}^{+\infty}
\int_{0}^{2\pi} \!\!\! \dots \int_{0}^{2\pi}
{\cal L} \left( \bm{\varphi}, \bm{\varphi}_{T}, \bm{\varphi}_{X},
\dots \right) \,\, {d^{m} \theta \over (2\pi)^{m}} \, dX \, dT
\end{equation}
where ${\cal L}$ is the Lagrangian density defined in 
(\ref{Lagrform}).

 After the substitution of (\ref{gradexp1}) in ${\cal L}$
(for known mapping $\tau$) we obtain the Lagrangian functional
depending on the functions $S^{\alpha}(X,T)$. Since the solutions
(\ref{gradexp1}) satisfy the system (\ref{Lf1}) the Lagrangian
functional thus defined has an extremal on the functions
$(S^{\alpha}(X,T))$ satisfying dynamical system (\ref{defsyst1}).
We have then a "Lagrangian form" for the flow (\ref{defsyst1}),
however, the form of the functional dependence of 
$\bm{\Phi}_{(k)}$ on the values ${\bf S}_{X}$, ${\bf S}_{T}$, 
${\bf S}_{XX}$, ${\bf S}_{XT}$, $\dots$, is supposed to be known 
in this approach. 

 The "averaged" Lagrangian density 
${\bar {\cal L}}[{\bf S}](X,T)$ contains an infinite number
of terms with the increasing numbers of $X$ (and $T$) 
differentiations. 

 The system (\ref{defsyst1}) can then be written in the
Lagrangian form

\begin{equation}
\label{deflagform}
{\delta \over \delta \, S^{\alpha}(X,T)} 
\int\int \sum_{k \geq 0} {\bar {\cal L}}_{(k)}
\left( {\bf S}_{X^{\prime}}, {\bf S}_{T^{\prime}}, \dots 
\right) \,\, d X^{\prime} \, dT^{\prime} \,\,\, = \,\,\, 0
\end{equation}
which gives an infinite formal expression containing the
higher derivatives of the function $S(X,T)$.

 Let us say, however, that the form (\ref{deflagform}) does
not coincide exactly with (\ref{defsyst1}) since it contains
the higher $T$-derivatives of ${\bf S}(X,T)$ like 
${\bf S}_{TTX}$, ${\bf S}_{TTXX}$, ${\bf S}_{TTTX}$,
${\bf S}_{TTXXX}$, $\dots$, in the formal expansion.
To get the system (\ref{defsyst1}) we have to resolve
the system (\ref{deflagform}) w.r.t. derivatives
$S^{\alpha}_{TT}$ and then to "remove recursively" all higher
$T$-derivatives ${\bf k}_{TT}$, ${\bm \omega}_{TT}$,
${\bf k}_{TTX}$, ${\bm \omega}_{TTX}$, ${\bf k}_{TTT}$, 
$\dots$, of the parameters ${\bf k}$ 
and $\bm{\omega}$ from the right-hand part. \footnote{Let us 
note that we assume that the function $S^{\alpha}_{mT,kX}$ 
($m \geq 2$) is represented by an infinite sum of terms having 
degrees $\geq m+k-1$.} After this procedure we will obtain
the infinite number of terms having certain degrees
in the right-hand part of our system which will coincide 
with the right-hand part of the system (\ref{defsyst1}).
We can say then that the procedure described above gives the
"Lagrangian formalism" for the system (\ref{defsyst1}).

\section{The deformation of the Whitham system for the 
non-linear "V-Gordon" equation.}
\setcounter{equation}{0}

 Let us consider the one-phase modulated solutions of 
the nonlinear "V-Gordon" equation

\begin{equation}
\label{VG1}
\varphi_{tt} \,\, - \,\, \varphi_{xx} \,\, + \,\,
V^{\prime} (\varphi) \,\,\, = \,\,\, 0 
\end{equation}

 The one-phase solutions of (\ref{VG1}) satisfy the equation

$$\left( S_{T}^{2} - S_{X}^{2} \right) \Phi_{\theta\theta} 
\, + \, V^{\prime} \left(\Phi\right) \,\, = \,\, 0 $$
which gives the well-known representation of one-phase
periodic solutions for the function $\Phi(\theta)$

$$\theta \, + \, \theta_{0} \,\,\, = \,\,\, 
\sqrt{\omega^{2} - k^{2}} \, \int
{d \, \Phi \over \sqrt{2(E - V(\Phi))}} $$ 
where the parameter $E(X,T)$ is connected with
$\omega(X,T)$, $k(X,T)$ by the formula

$$\oint {d \, \Phi \over \sqrt{2(E - V(\Phi))}}
\,\,\, = \,\,\, {2\pi \over \sqrt{\omega^{2} - k^{2}}} $$

\begin{figure}
\begin{center}
\epsfig{file=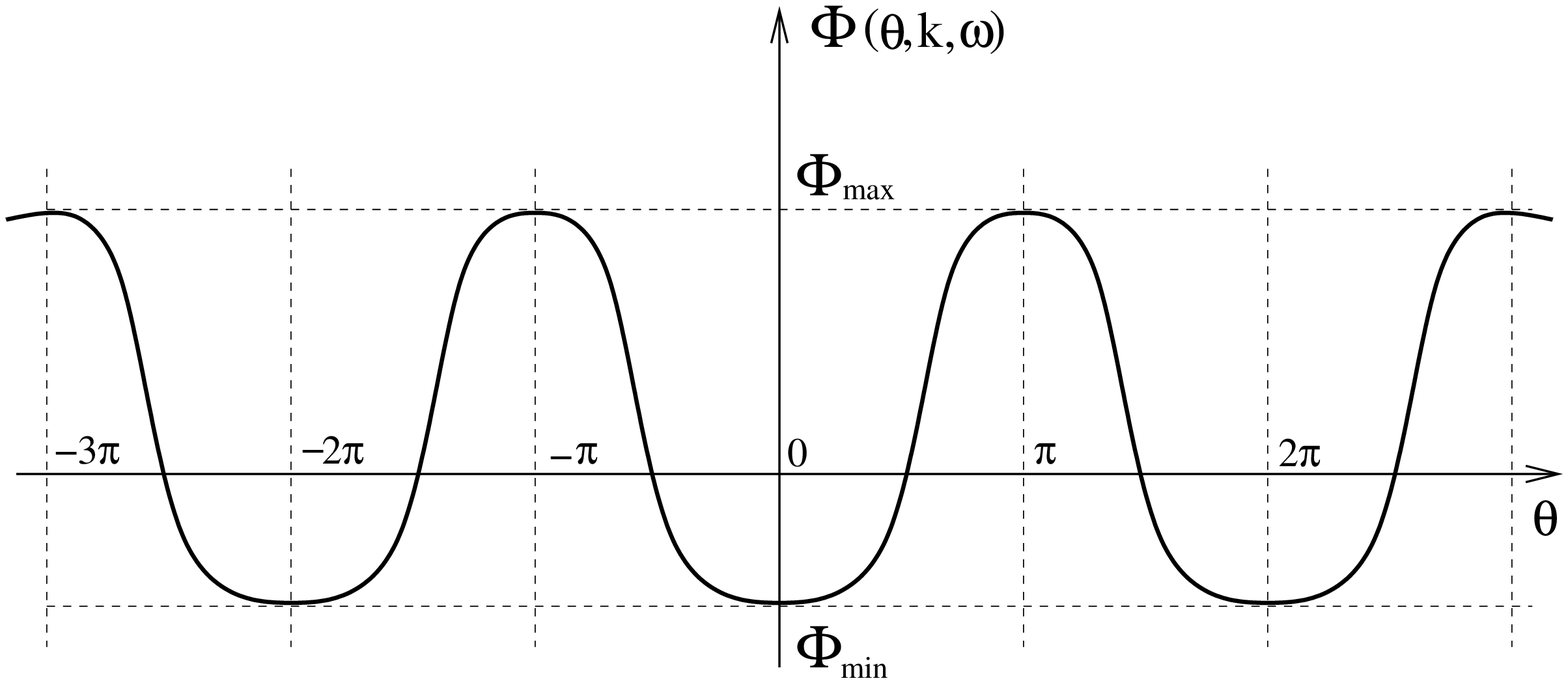,width=14.0cm,height=7cm}
\end{center}
\caption{The function $\Phi(\theta,k,\omega)$ having zero initial 
phase shift.}
\label{zerophase}
\end{figure}

 Let us choose now the functions $\Phi(\theta,k,\omega)$
(having zero initial phase shifts) such that
$\Phi(\theta,k,\omega)$ has a local minimum at the point
$\theta = 0$ for all $k$ and $\omega$ (Fig. \ref{zerophase}).

 We take now

$$\Phi_{(0)}(\theta, X, T) \,\, = \,\,
\Phi (\theta, S_{X}, S_{T}) $$
for the zero approximation in (\ref{gradexp1}).

 Let us say that the functions $S_{TT}$, $\Phi_{\theta T}$,
$\Phi_{TT}$ do not have certain degrees in our approach since the
time derivatives $\omega_{T}$, $\omega_{TT}$ are represented by  
the infinite series given by (\ref{defsyst1}). However, we can   
claim that the functions $S_{TT}$, $\Phi_{\theta T}$ are given   
by the sums of terms having degrees $\geq 1$. In the same way    
the function $\Phi_{TT}$ is represented by the sum of terms all  
having degrees $\geq 2$.

 Let us use now the following notation: 

 Namely, we will denote by the symbols like $S_{TT}^{[k]}$,
$\Phi_{\theta T}^{[k]}$, $\Phi_{TT}^{[k]}$, $\dots$, the terms
of the degree $k$ in the infinite expansions of corresponding
expressions. 

 We can write then

$$\left( S_{T}^{2} - S_{X}^{2} \right) \Phi_{(1)\theta\theta}
\, + \, V^{\prime\prime} \left(\Phi_{(0)}\right)  \Phi_{(1)}
\,\, = \,\, \tilde{f}_{(1)}(\theta, X, T) $$
where

\begin{equation}
\label{firstdiscr}
\tilde{f}_{(1)}(\theta, X, T) 
\,\, = \,\, 2 S_{X} \, \Phi_{(0)\theta X}
\, - \, 2 S_{T} \, \Phi^{[1]}_{(0)\theta T} \, + \,
\left( S_{XX} - S^{[1]}_{TT} \right)  \Phi_{(0)\theta} 
\end{equation}
for the first approximation in (\ref{gradexp1}).

 The orthogonality condition

$$\int_{0}^{2\pi} \tilde{f}_{(1)} \, \Phi_{(0)\theta} \, 
{d \theta \over 2\pi} \,\, = \,\, 0 $$
gives the first term (Whitham system) of the system (\ref{defsyst1})
having the form

$$\left( S^{[1]}_{TT} - S_{XX} \right) \int_{0}^{2\pi} 
\Phi_{(0)\theta}^{2} {d \theta \over 2\pi} \, + \,
S_{T} \left[ \int_{0}^{2\pi} \Phi_{(0)\theta}^{2} 
{d \theta \over 2\pi} \right]_{T}^{[1]} \, - \,
S_{X} \left[ \int_{0}^{2\pi} \Phi_{(0)\theta}^{2}
{d \theta \over 2\pi} \right]_{X} \,\, = \,\, 0 $$
or, equivalently

\begin{equation}
\label{ws1}
\left[ S_{T} \int_{0}^{2\pi} \Phi_{(0)\theta}^{2}
{d \theta \over 2\pi} \right]_{T}^{[1]} \,\,\, = \,\,\,
\left[ S_{X} \int_{0}^{2\pi} \Phi_{(0)\theta}^{2}
{d \theta \over 2\pi} \right]_{X}
\end{equation}

 We have then

$$\left[ S_{T} \int_{0}^{2\pi} \Phi_{(0)\theta}^{2}
{d \theta \over 2\pi} \right]_{S_{T}} S_{TT}^{[1]}
\, + \, \left[ S_{T} \int_{0}^{2\pi} \Phi_{(0)\theta}^{2}
{d \theta \over 2\pi} \right]_{S_{X}}  S_{XT}
\,\,\, = \,\,\, \left[ S_{X} \int_{0}^{2\pi} \Phi_{(0)\theta}^{2}
{d \theta \over 2\pi} \right]_{X} $$
where $S_{TT}^{[1]} = \sigma_{(1)} (S_{X},S_{T},S_{XX},S_{XT})$.

 Finally we get

\begin{equation} 
\label{sigma1}
\sigma_{(1)} \,\,\, = \,\,\, \left(
\left[ S_{X} \int_{0}^{2\pi} \Phi_{(0)\theta}^{2}
{d \theta \over 2\pi} \right]_{X} \, - \, 
\left[ S_{T} \int_{0}^{2\pi} \Phi_{(0)\theta}^{2}
{d \theta \over 2\pi} \right]_{S_{X}}  S_{XT} \right)
\left/  
\left[ S_{T} \int_{0}^{2\pi} \Phi_{(0)\theta}^{2}
{d \theta \over 2\pi} \right]_{S_{T}} \right.
\end{equation}
where

$$\int_{0}^{2\pi} \Phi_{(0)\theta}^{2} {d \theta \over 2\pi}
\,\,\, = \,\,\, \oint 
{\sqrt{2(E-V(\Phi))} \over \sqrt{\omega^{2} - k^{2}}}
\,\, {d \Phi \over 2\pi} $$

 The higher systems (\ref{newksyst}) can be written in 
analogous form

\begin{equation}
\label{VGksyst}
\left( S_{T}^{2} - S_{X}^{2} \right) \Phi_{(k)\theta\theta}
\, + \, V^{\prime\prime} \left(\Phi_{(0)}\right)  \Phi_{(k)}
\,\, = \,\, \tilde{f}_{(k)}(\theta, X, T)
\end{equation}
where the orthogonality conditions

\begin{equation}
\label{VGortcond}
\int_{0}^{2\pi} \tilde{f}_{(k)} \, \Phi_{(0)\theta} \, 
{d \theta \over 2\pi} \,\,\, = \,\,\, 0
\end{equation}
are imposed for all $k \geq 1$.

 We impose also the normalization conditions

\begin{equation}
\label{VGnorm}
\int_{0}^{2\pi} \Phi_{(k)} \, \Phi_{(0)\theta} \,
{d \theta \over 2\pi} \,\,\, = \,\,\, 0
\end{equation}
for all $k \geq 1$.

 Let us look for a solution of (\ref{VGksyst}) in the form
(see also \cite{luke,AblBenny})

$$\Phi_{(k)}(\theta, X, T) \,\, = \,\, 
\alpha_{(k)}(\theta, X, T) \, \Phi_{(0)\theta}(\theta, X, T) $$
where $\alpha_{(k)}(\theta, X, T)$ is the function 
$2\pi$-periodic in $\theta$. We have

$$\left( S_{T}^{2} - S_{X}^{2} \right) \alpha_{(k)\theta\theta}
\, \Phi_{(0)\theta} \,\, + \,\, 
2 \left( S_{T}^{2} - S_{X}^{2} \right) \alpha_{(k)\theta} \,
\Phi_{(0)\theta\theta} \,\, = \,\, {\tilde f}_{(k)} $$
or

$$\left( S_{T}^{2} - S_{X}^{2} \right) \alpha_{(k)\theta}
\left( \Phi_{(0)\theta} \right)^{2} \,\,\, = \,\,\,
\int^{\theta} \Phi_{(0)\theta^{\prime}} \,
{\tilde f}_{(k)}(\theta^{\prime}) \, d \theta^{\prime} 
\,\, + \,\, \xi_{1} $$
where $\xi_{1}$ is arbitrary constant.

 We have then

\begin{equation}
\label{Phik}
\Phi_{(k)} \,\, = \,\, 
{\Phi_{(0)\theta} \over S_{T}^{2} - S_{X}^{2}} \int^{\theta} 
{d \theta^{\prime} \over ( \Phi_{(0)\theta^{\prime}} )^{2} } 
\int^{\theta^{\prime}} \Phi_{(0)\theta^{\prime\prime}} \,
{\tilde f}_{(k)}(\theta^{\prime\prime}) \, 
d \theta^{\prime\prime} \,\, + \,\, 
{\xi_{1} \, \Phi_{(0)\theta} \over S_{T}^{2} - S_{X}^{2}}
\int^{\theta}
{d \theta^{\prime} \over ( \Phi_{(0)\theta^{\prime}} )^{2} }
\,\, + \,\, 
{\xi_{2} \, \Phi_{(0)\theta} \over S_{T}^{2} - S_{X}^{2}}
\end{equation}

 However, the formula (\ref{Phik}) has a local character and
we have to investigate the solution (\ref{Phik}) on the whole
axis $- \infty < \theta < + \infty $. Let us remind that we
assume that the conditions (\ref{VGortcond}) are satisfied.
Easy to see that the expression $1/(\Phi_{(0)\theta})^{2}$
has singularities at the points $\theta_{n} = \pi n$,
$n \in \mathbb{Z}$, so the integration should be made
carefully in the formula (\ref{Phik}). Let us consider now
two important cases arising:

I) The function ${\tilde f}^{(k)}(\theta)$ is anti-symmetric
in $\theta$:  
${\tilde f}^{(k)}(-\theta) = - {\tilde f}^{(k)}(\theta)$;

II) The function ${\tilde f}^{(k)}(\theta)$ is symmetric 
in $\theta$:
${\tilde f}^{(k)}(-\theta) = {\tilde f}^{(k)}(\theta)$.

 Let us start with the case (I). In the case (I) we have in 
fact from (\ref{VGortcond})

$$\int_{0}^{\pi} \tilde{f}_{(k)} \, \Phi_{(0)\theta} \,
{d \theta \over 2\pi} \,\,\, = \,\,\, 
\int_{-\pi}^{0} \tilde{f}_{(k)} \, \Phi_{(0)\theta} \,
{d \theta \over 2\pi} \,\,\, = \,\,\,  
\int_{\pi n}^{\pi(n+1)} \tilde{f}_{(k)} \, \Phi_{(0)\theta} \,
{d \theta \over 2\pi} \,\,\, = \,\,\, 0 $$
view the anti-symmetry (and periodicity) of the function
$\Phi_{(0)}(\theta)$.

 It is easy to see then that the expression
 
\begin{equation}
\label{expr1}
{1 \over ( \Phi_{(0)\theta^{\prime}} )^{2} }
\int_{0}^{\theta^{\prime}} \Phi_{(0)\theta^{\prime\prime}} \,   
{\tilde f}_{(k)}(\theta^{\prime\prime}) \,
d \theta^{\prime\prime}
\end{equation}
has in fact no singularities at the points 
$\theta^{\prime}_{n} = \pi n$. Moreover, the expression
(\ref{expr1}) defines an anti-symmetric periodic function
of $\theta^{\prime}$ so we have

$$\int_{0}^{2\pi} 
{d \theta^{\prime} \over ( \Phi_{(0)\theta^{\prime}} )^{2} }
\int_{0}^{\theta^{\prime}} \Phi_{(0)\theta^{\prime\prime}} \,   
{\tilde f}_{(k)}(\theta^{\prime\prime}) \,
d \theta^{\prime\prime} \,\,\, = \,\,\, 0$$

 We can see then that the expression

$${\Phi_{(0)\theta} \over S_{T}^{2} - S_{X}^{2}} 
\int_{0}^{\theta}
{d \theta^{\prime} \over ( \Phi_{(0)\theta^{\prime}} )^{2} }
\int_{0}^{\theta^{\prime}} \Phi_{(0)\theta^{\prime\prime}} \,   
{\tilde f}_{(k)}(\theta^{\prime\prime}) \,
d \theta^{\prime\prime} $$
gives a smooth periodic anti-symmetric solution of (\ref{VGksyst}).
We can omit now the local parameters $\xi_{1}$, $\xi_{2}$
and use the global freedom in the periodic solution 
of (\ref{VGksyst}) defined modulo the (anti-symmetric)
function $a_{(k)}(X,T) \Phi_{(0)\theta}(\theta,X,T)$ 
to satisfy the normalization condition (\ref{VGnorm}). 
We can formulate now the following Proposition:

\vspace{0.5cm}

{\bf Proposition 3.1.}

{\it For a smooth periodic anti-symmetric discrepancy function
${\tilde f}_{(k)}(\theta)$ the solution 
$\Phi_{(k)}(\theta)$ of (\ref{VGksyst}) satisfying the 
normalization conditions (\ref{VGnorm}) is a smooth periodic
anti-symmetric function 
$\Phi_{(k)}(-\theta) = - \Phi_{(k)}(\theta)$. }

\vspace{0.5cm}

 Let us consider now the case (II). 

\begin{figure}
\begin{center}
\epsfig{file=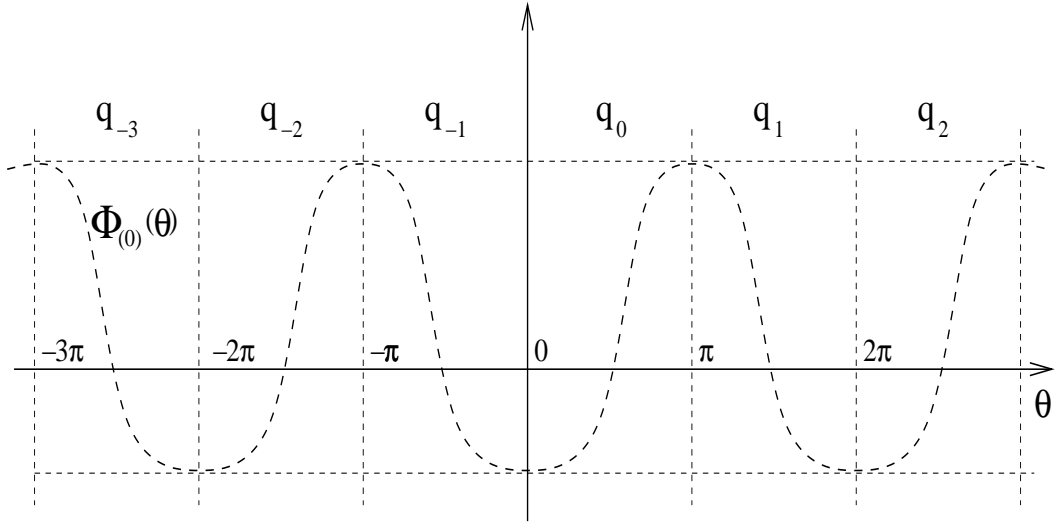,width=14.0cm,height=7cm}
\end{center}
\caption{The "odd" and "even" intervals $q_{n}$ on the axis
$-\infty < \theta < +\infty$.}
\label{symcase}
\end{figure}

 Let us divide the axis $-\infty < \theta < +\infty$ into
"even" ($2\pi l < \theta < 2\pi l + \pi$) and "odd"
($2\pi l - \pi < \theta < 2\pi l$) intervals $q_{n}$
(Fig. \ref{symcase}).

 Let us define the "piecewise" solution $\Phi_{(k)}$ defined
on any interval $q_{n}$ by the formula

$$\Phi_{(k),\{n\}} \, = \, 
{\Phi_{(0)\theta} \over S_{T}^{2} - S_{X}^{2}} 
\int_{\pi n + \pi/2}^{\theta}
{d \theta^{\prime} \over ( \Phi_{(0)\theta^{\prime}} )^{2} }
\int_{\pi n + \pi/2}^{\theta^{\prime}} 
\Phi_{(0)\theta^{\prime\prime}} \,   
{\tilde f}_{(k)}(\theta^{\prime\prime}) \,
d \theta^{\prime\prime} \, + $$
$$+ \,
{\xi_{1,\{n\}} \, \Phi_{(0)\theta} \over S_{T}^{2} - S_{X}^{2}}
\int_{\pi n + \pi/2}^{\theta}
{d \theta^{\prime} \over ( \Phi_{(0)\theta^{\prime}} )^{2} }
\, + \,
{\xi_{2,\{n\}} \, \Phi_{(0)\theta} \over S_{T}^{2} - S_{X}^{2}}$$
where the coefficients $\xi_{1,\{n\}}$, $\xi_{2,\{n\}}$ depend on 
the interval $q_{n}$. Let us put now

$$\xi_{1,\{2l\}} \, = \, \xi_{1,\{2l+1\}} \, = \, \xi_{1} 
\,\,\, , \,\,\,\,\, 
\xi_{2,\{2l\}} \, = \, - \xi_{2,\{2l+1\}} \, = \, \xi_{2}
\,\,\, , \,\,\,\,\, l \in \mathbb{Z} $$

 It's not difficult to see then that the total solution 
$\Phi_{(k)}$ given by all $\Phi_{(k),\{n\}}$ is a symmetric
continuous periodic function of $\theta$. Using the parameters
$\xi_{1}$, $\xi_{2}$ we can provide also the $C_{1}$ 
smoothness both on "odd" ($\theta = 2\pi l + \pi$) and "even"
($\theta = 2\pi l$) sides of the intervals $q_{n}$. It appears
then that the function $\Phi_{(k)}$ is smooth for the smooth
${\tilde f}_{(k)}$ since it satisfies the second-order
equation (\ref{VGksyst}). The conditions (\ref{VGnorm})
are automatically satisfied here view the symmetry of the 
function $\Phi_{(k)}$. We can formulate the Proposition:

\vspace{0.5cm}

{\bf Proposition 3.2.}

{\it For a smooth periodic symmetric discrepancy function
${\tilde f}_{(k)}(\theta)$ the solution
$\Phi_{(k)}(\theta)$ of (\ref{VGksyst}) satisfying the
normalization conditions (\ref{VGnorm}) is a smooth periodic
symmetric function
$\Phi_{(k)}(-\theta) = \Phi_{(k)}(\theta)$. }
 
\vspace{0.5cm}

 Thus we have for the first correction $\Phi_{(1)}$
in (\ref{gradexp1}):

$${\tilde f}_{(1)}(\theta, X, T) 
\, = \, 2 S_{X} \Phi_{(0)\theta X} \, - \,
2 S_{T} \left[ \Phi_{(0)\theta T} \right]^{[1]} \, + \, 
\left( S_{XX} - S_{TT}^{[1]} \right) \Phi_{(0)\theta} \, = $$

$$= \, 2 k \Phi_{(0)\theta k} k_{X} \, + \, 
2 k \Phi_{(0)\theta \omega} \omega_{X} \, - \, 
2 S_{T} \Phi_{(0)\theta k} \omega_{X} \, - $$

\begin{equation}
\label{f1expr}
- \, 2 S_{T} \Phi_{(0)\theta \omega} 
\sigma_{(1)}(k,\omega,k_{X},\omega_{X}) \, + \,
\Phi_{(0)\theta} k_{X} \, - \, \Phi_{(0)\theta}
\sigma_{(1)}(k,\omega,k_{X},\omega_{X}) 
\end{equation}
where the function $\sigma_{(1)}$ is defined by (\ref{sigma1}).

 Easy to see that the function ${\tilde f}_{(1)}$ is 
anti-symmetric 
${\tilde f}_{(1)}(-\theta) = - {\tilde f}_{(1)}(\theta)$.
We obtain then that the function $\Phi_{(1)}(\theta)$
is also anti-symmetric in $\theta$:
$\Phi_{(1)}(-\theta) = - \Phi_{(1)}(\theta)$.

 We have then for $\Phi_{(2)}(\theta,X,T)$

$$\left( S_{T}^{2} - S_{X}^{2} \right) \Phi_{(2)\theta\theta}
\, + \, V^{\prime\prime} \left(\Phi_{(0)}\right)  \Phi_{(2)}
\,\, = \,\, \tilde{f}_{(2)}(\theta, X, T) $$
where 

$$\tilde{f}_{(2)}(\theta, X, T) \,\, = \,\, 
- \, 2 S_{T} \left[ \Phi_{(0)\theta T} \right]^{[2]} 
\, - \, S_{TT}^{[2]} \Phi_{(0)\theta}
\, - \, S_{TT}^{[1]} \Phi_{(1)\theta} \, - \,
{1 \over 2} V^{\prime\prime\prime} \left( \Phi_{(0)} \right)
\Phi_{(1)}^{2} \, + \,
2 S_{X} \Phi_{(1)\theta X} \, - $$
$$- \, 2 S_{T} \left[ \Phi_{(1)\theta T} \right]^{[2]} 
\, + \, S_{XX} \Phi_{(1)\theta}
\, - \, \left[ \Phi_{(0)TT} \right]^{[2]} \, + \,
\Phi_{(0)XX} \,\, = $$

\vspace{0.5cm}

$$= \,\, - \left( 2 \omega \Phi_{(0)\theta\omega} \, + \,
 \Phi_{(0)\theta} \right) \sigma_{(2)}
(k, \omega, k_{X}, \omega_{X}, k_{XX}, \omega_{XX}) \,\, + $$
$$- \, \Phi_{(1)\theta}
\sigma_{(1)} (k, \omega, k_{X}, \omega_{X}) \, - \,
{1 \over 2} V^{\prime\prime\prime} \left( \Phi_{(0)} \right)
\Phi_{(1)}^{2} + \, 2 k \Phi_{(1)\theta X} \, - $$
$$- \, 2 \omega 
\int {\delta \Phi_{(1)\theta} (\theta,X) \over \delta k(Z)}
\, \omega_{Z} \, dZ \,\, - \,\, 2 \omega \int 
{\delta \Phi_{(1)\theta} (\theta,X) \over \delta \omega (Z)} \,
\sigma_{(1)} (k, \omega, k_{Z}, \omega_{Z}) \, dZ \, + \, 
k_{X} \Phi_{(1)\theta} \, - $$ 
$$- \,\, \Phi_{(0)kk}
(\omega_{X})^{2} \, - \, 2 \Phi_{(0)k\omega} \, \omega_{X} \,
\sigma_{(1)} (k, \omega, k_{X}, \omega_{X}) \, - \,
\Phi_{(0)k} \, {d \over dX} 
\sigma_{(1)} (k, \omega, k_{X}, \omega_{X}) \,\, - $$
$$- \,\, \Phi_{(0)\omega\omega} \, 
\sigma_{(1)}^{2} (k, \omega, k_{X}, \omega_{X}) \, - \,
\Phi_{(0)\omega} \int
{\delta \sigma_{(1)} (k, \omega, k_{X}, \omega_{X}) \over
\delta k(Z)} \, \omega_{Z} \, dZ \,\, - $$
$$- \,\, \Phi_{(0)\omega} \int
{\delta \sigma_{(1)} (k, \omega, k_{X}, \omega_{X}) \over
\delta \omega(Z)} 
\, \sigma_{(1)} (k, \omega, k_{Z}, \omega_{Z})
\, dZ \,\,\, + \,\,\, \Phi_{(0)XX} $$

 Easy to see that only the first two terms in ${\tilde f}_{(2)}$
(containing $\sigma_{(2)}$) are anti-symmetric in $\theta$
and all the other terms are symmetric. Using the orthogonality
conditions (\ref{VGortcond}) we obtain then

\begin{equation}
\label{sigma2}
\sigma_{(2)}(k, \omega, k_{X}, \omega_{X}, k_{XX}, \omega_{XX})
\,\, \equiv \,\, 0
\end{equation}
for the second term of the deformation of Whitham system
(\ref{defsyst1}).

 Using (\ref{sigma2}) we see then that the discrepancy 
function ${\tilde f}_{(2)}$ becomes now a symmetric function 
of $\theta$, so we can state that the second approximation
$\Phi_{(2)}(\theta, X, T)$ in (\ref{gradexp1}) is a 
symmetric function of $\theta$: 
$\Phi_{(2)}(-\theta) = \Phi_{(2)}(\theta)$.
Using the simple induction we can claim in fact that all
the discrepancy functions ${\tilde f}_{(k)}$, $k \geq 1$
can be represented in the form

$${\tilde f}_{(k)}(\theta, X, T) \,\, = \,\, - \,
\left( 2 \omega \Phi_{(0)\theta\omega} \, + \,
 \Phi_{(0)\theta} \right) \sigma_{(k)}
(k, \omega, \dots) \,\, + \,\, 
{\tilde f}_{(k)}^{\prime}(\theta, X, T) $$
where ${\tilde f}_{(k)}^{\prime}$ does not contain the 
function $\sigma_{(k)}(k, \omega, \dots)$. Moreover,
all the "even" functions ${\tilde f}_{(2l)}^{\prime}$
will be symmetric in $\theta$ and all the "odd" functions
${\tilde f}_{(2l+1)}^{\prime}$ will be anti-symmetric
in $\theta$. The same is true also for the functions
${\tilde f}_{(2l)}$, ${\tilde f}_{(2l+1)}$ after the
imposing of the conditions (\ref{VGortcond}) and for the
corresponding solutions of (\ref{VGksyst}) 
$\Phi_{(2l)}$, $\Phi_{(2l+1)}$.\footnote{The similar facts
were pointed out in \cite{AblBenny} for "V"-Gordon equation.
However the functions $\Phi_{(k)}$, ${\tilde f}_{(k)}$
are different here from those appeared in \cite{AblBenny}.}
We can formulate then the following Lemma:

\vspace{0.5cm}

{\bf Lemma 3.1.}

{\it For the "unified" choice of the functions
$\Phi(\theta, k, \omega)$ corresponding to 
Fig. \ref{zerophase} the following statements are true:

1) All the even terms $\sigma_{(2l)}(k, \omega, \dots)$
in the deformation of Whitham system (\ref{defsyst1}) are
identically zero: $\sigma_{(2l)} \equiv 0$;

2) All the odd corrections $\Phi_{(2l+1)}(\theta,X,T)$,
$l \geq 0$ in (\ref{gradexp1}) are anti-symmetric in $\theta$:
$\Phi_{(2l+1)}(-\theta) = - \Phi_{(2l+1)}(\theta)$;

3) All the even corrections $\Phi_{(2l)}(\theta,X,T)$,
$l \geq 1$ in (\ref{gradexp1}) are symmetric in $\theta$:
$\Phi_{(2l)}(-\theta) = \Phi_{(2l)}(\theta)$. }

\vspace{0.5cm}

 The system (\ref{defsyst1}) for the given choice of 
functions $\Phi(\theta, k, \omega)$ can be rewritten 
now in the form

$$k_{T} \,\,\, = \,\,\, \omega_{X} $$
\begin{equation}
\label{defsyst2}
\omega_{T} \,\,\, = \,\,\, \sum_{l \geq 0}
\sigma_{(2l+1)} (k, \omega, k_{X}, \omega_{X}, \dots)
\end{equation} 
where all $\sigma_{(2l+1)}$ are polynomial in derivatives
$k_{X}$, $\omega_{X}$, $\dots$, having degree $(2l + 1)$.

 Let us say here that the form of deformations of
Hydrodynamic hierarchies containing only odd dispersion
terms was considered first by B.A. Dubrovin and Y. Zhang
(\cite{DubrZhang1,DubrZhang2})
as the convenient representative in the class of 
"equivalent" deformations.

\vspace{0.5cm}

 Finally, let us calculate the next ($\sigma_{3}$) 
non-vanishing term in (\ref{defsyst2}). It is convenient to
use now the Lagrangian formalism of the system (\ref{VG1})

$$\delta \int\int \left[ - {1 \over 2} \varphi_{t}^{2} 
\, + \, {1 \over 2} \varphi_{x}^{2} \, + \, V(\varphi) \right] 
\, dx \, dt \,\,\, = \,\,\, 0 $$

 We have to add the variable $\theta$ and introduce the 
action functional

\begin{equation}
\label{lagrVG}  
\Sigma [\varphi] \,\,\, = \,\,\, \int\int \int_{0}^{2\pi}
\left[ - {1 \over 2} \varphi_{T}^{2} \, + \,
{1 \over 2} \varphi_{X}^{2} \, + \, V(\varphi) \right] \,
{d \theta \over 2\pi} \, dX \, dT 
\end{equation}

 Let us introduce the function $\Phi^{(tot)}(\theta,X,T)$
by the formula

$$\Phi^{(tot)}(\theta,X,T) \,\,\, = \,\,\,
\sum_{k \geq 0} \Phi_{(k)} (\theta,X,T) \,\,\, = \,\,\,
\phi(\theta - S(X,T), X, T) $$
(where $\Phi_{(0)} (\theta,X,T) = \Phi(\theta, S_{X}, S_{T})$).

 After the substitution of (\ref{gradexp1}) into the functional
(\ref{lagrVG}) we can write the action functional in the form

$$\Sigma \,\,\, = \,\,\, \int\int \int_{0}^{2\pi}
\left[ - \, {1 \over 2} \, S_{T}^{2} 
\left( \Phi^{(tot)}_{\theta} \right)^{2} \, + \,
{1 \over 2} \, 
S_{X}^{2} \left( \Phi^{(tot)}_{\theta} \right)^{2} \, + \,
V \left(\Phi^{(tot)} \right) \right] \, 
{d \theta \over 2\pi} \, dX \, dT \,\, + $$   
$$+ \,\, \int\int \int_{0}^{2\pi} \left[ - \, 
S_{T} \Phi^{(tot)}_{\theta} \Phi^{(tot)}_{T} \, + \,
S_{X} \Phi^{(tot)}_{\theta} \Phi^{(tot)}_{X} \right] \,
{d \theta \over 2\pi} \, dX \, dT \,\, + $$
$$ + \,\, \int\int \int_{0}^{2\pi} {1 \over 2} \left[ - \,
\left( \Phi^{(tot)}_{T} \right)^{2} \, + \,
\left( \Phi^{(tot)}_{X} \right)^{2} \right] \,
{d \theta \over 2\pi} \, dX \, dT \,\,\,  = $$

$$= \,\,\, \int\int \int_{0}^{2\pi} \left(
{\cal L}^{\prime} \, + \, {\cal L}^{\prime\prime} \, + \,
{\cal L}^{\prime\prime\prime} \right) \,
{d \theta \over 2\pi} \, dX \, dT $$

 The function

$${\bar {\cal L}} [S] \,\,\, = \,\,\, \int_{0}^{2\pi}
\left( {\cal L}^{\prime}[S] \, + \, 
{\cal L}^{\prime\prime}[S] \, + \,
{\cal L}^{\prime\prime\prime}[S] \right) \,
{d \theta \over 2\pi} $$
gives a Lagrangian density for the "averaged" Lagrangian
formalism corresponding to the system (\ref{defsyst2}).

 The main term ${\bar {\cal L}}_{(0)}$ of the averaged Lagrangian 
density ${\bar {\cal L}}$ has the form

$${\bar {\cal L}}_{(0)} (S_{X}, S_{T}) \,\,\, = \,\,\,
\int_{0}^{2\pi}
\left[ {1 \over 2} \left( - \, S_{T}^{2} \, + \,
S_{X}^{2} \right) \Phi_{(0)\theta}^{2} \, + \,
V \left( \Phi_{(0)} \right) \right] \,
{d \theta \over 2\pi} $$

 The equation

$$\left[ {\delta \over \delta S(X,T)} \int\int
{\bar {\cal L}}_{(0)} (S_{X}, S_{T}) \, dX \, dT 
\right]^{[1]} \,\,\, = \,\,\, 0 $$
gives the Whitham's Lagrangian formalism for the
Whitham system (\ref{ws1}). (We use as previously the
notation $^{[1]}$ for the terms of degree $1$ in the
"gradated" expansion of corresponding expression).

 For the determination of the first non-trivial term 
$\sigma_{(3)}$ in the deformation
(\ref{defsyst2}) it is enough to calculate the correction 
${\bar {\cal L}}_{(2)}$ to the main term ${\bar {\cal L}}_{(0)}$ 
given by the expression

$$\int\!\dots\!\int\!\!
\int_{0}^{2\pi} \!\!\!\!\! \dots \! 
\int_{0}^{2\pi} {1 \over 2} \,
{ \delta^{2} {\cal L}^{\prime}(\theta, X, T) \over
\delta \Phi (\theta^{\prime}, X^{\prime}, T^{\prime}) \,
\delta \Phi (\theta^{\prime\prime}, X^{\prime\prime}, 
T^{\prime\prime}) } |_{\Phi_{(0)}} \,
\Phi_{(1)} (\theta^{\prime}, X^{\prime}, T^{\prime}) \,
\Phi_{(1)} (\theta^{\prime\prime}, X^{\prime\prime},
T^{\prime\prime}) \,\, \times $$
$$\hspace{9cm} \times \,
d\theta^{\prime} dX^{\prime} dT^{\prime} \,
d\theta^{\prime\prime} dX^{\prime\prime} dT^{\prime\prime} \,\,
{d \theta \over 2\pi} \,\,\, + $$

$$+ \,\, \int\dots\int 
\int_{0}^{2\pi} \!\!\! \dots \int_{0}^{2\pi} 
{ \delta {\cal L}^{\prime\prime}(\theta, X, T) \over
\delta \Phi (\theta^{\prime}, X^{\prime}, T^{\prime}) }
|_{\Phi_{(0)}} \,
\Phi_{(1)} (\theta^{\prime}, X^{\prime}, T^{\prime}) 
\, d\theta^{\prime} dX^{\prime} dT^{\prime} \,\,
{d \theta \over 2\pi} \,\,\, + $$ 

$$+ \,\, \int_{0}^{2\pi} 
{\cal L}^{\prime\prime\prime}(\theta, X, T)|_{\Phi_{(0)}}
\, {d \theta \over 2\pi} $$

 Using the normalization condition (\ref{VGnorm}) we can write
the correction $\Phi_{(1)}(\theta,X,T)$ as a local functional of
$(S_{X}, S_{T}, S_{XX}, S_{XT})$ having the form

$$\Phi_{(1)}(\theta,X,T) \,\, = \,\,
{\Phi_{\theta} \over S_{T}^{2} - S_{X}^{2}} \int_{0}^{\theta} 
{d \theta^{\prime} \over (\Phi_{\theta^{\prime}})^{2}}
\int_{0}^{\theta^{\prime}} \Phi_{\theta^{\prime\prime}}
{\tilde f}_{(1)} (\theta^{\prime\prime}) \,
d \theta^{\prime\prime} \,\, - $$
$$- \,\, \Phi_{\theta} \int_{0}^{2\pi}
\Phi_{\theta^{\prime}}^{2} \, d \theta^{\prime}
\int_{0}^{\theta^{\prime}} 
{d \theta^{\prime\prime} \over (\Phi_{\theta^{\prime\prime}})^{2}}
\int_{0}^{\theta^{\prime\prime}}
\Phi_{\theta^{\prime\prime\prime}}
{\tilde f}_{(1)} (\theta^{\prime\prime\prime}) \,
d \theta^{\prime\prime\prime} \left/
\left(S_{T}^{2} - S_{X}^{2}\right) \int_{0}^{2\pi}
\Phi_{\theta^{\prime}}^{2} \, d \theta^{\prime} \right. $$
where ${\tilde f}_{(1)}(\theta,X,T)$ is given by formula
(\ref{f1expr}).

 We have then

$$\Sigma_{(2)} \, = \, {1 \over 2}
\int\!\!\int\!\!\int_{0}^{2\pi} \!
\left( \left[- S_{T}^{2} + S_{X}^{2} \right] \, 
\Phi_{(1)\theta}^{2} \, + \, V^{\prime\prime}(\Phi) \, 
\Phi_{(1)}^{2} \right) 
\,\, {d \theta \over 2\pi} \, dX \, dT \,\, +$$
$$+ \,\, \int\!\!\int\!\!\int_{0}^{2\pi} \!
\left(S_{TT} \Phi_{\theta} \, + \, 2 S_{T} \, \Phi_{\theta T}
\, - \, S_{XX} \Phi_{\theta} \, - \, 2 S_{X} \, \Phi_{\theta X}
\right) \Phi_{(1)}
\,\, {d \theta \over 2\pi} \, dX \, dT \,\, -$$
$$- \,\, {1 \over 2} \int\!\!\int\!\!\int_{0}^{2\pi}
\left( \Phi_{k}(\theta, S_{X}, S_{T}) \, S_{TX} \, + \,
\Phi_{\omega}(\theta, S_{X}, S_{T}) \, S_{TT} \right)^{2} \,
{d \theta \over 2\pi} \, dX \, dT \,\, +$$ 
$$+ \,\, {1 \over 2} \int\!\!\int\!\!\int_{0}^{2\pi}
\left( \Phi_{k}(\theta, S_{X}, S_{T}) \, S_{XX} \, + \,
\Phi_{\omega}(\theta, S_{X}, S_{T}) \,  S_{TX} \right)^{2} \,
{d \theta \over 2\pi} \, dX \, dT $$

 Let us consider now the equations

$$\left[ {\delta \over \delta S(X,T)} \,\, \left(
\Sigma_{(0)} [S] \,\, + \,\, \Sigma_{(2)} [S] \right)
\right]^{[1]} \,\,\, = \,\,\, 0 $$
and

$$\left[ {\delta \over \delta S(X,T)} \,\, \left(
\Sigma_{(0)} [S] \,\, + \,\, \Sigma_{(2)} [S] \right)
\right]^{[3]} \,\,\, = \,\,\, 0 $$

 The first equation gives again the Whitham system (\ref{ws1}).
The second equation defines the first non-trivial correction
$\sigma_{(3)}$ of the system (\ref{defsyst2}). We have

$$\left[ {\delta \over \delta S(X,T)} \,\, 
\Sigma_{(0)} [S] \right]^{[3]} \,\,\, = \,\,\,
\left[ S_{T} \int_{0}^{2\pi} \!
\Phi_{\theta}^{2} 
\,\, {d \theta \over 2\pi} \right]_{S_{T}} \,\,
S_{TT}^{[3]} $$
where $S_{TT}^{[3]} = \sigma_{(3)} (S_{X}, S_{T}, S_{XX}, 
S_{XT}, S_{XXX}, S_{XXT}, S_{XXXX}, S_{XXXT})$.

 Now finally we obtain

$$\sigma_{(3)} \,\,\, = \,\,\, - \,\, 
\left[ {\delta \over \delta S(X,T)} \,\,
\Sigma_{(2)} [S] \right]^{[3]} 
\left/
\left[ S_{T} \int_{0}^{2\pi} \!
\Phi_{\theta}^{2} \,\,
{d \theta \over 2\pi} \right]_{S_{T}} \right. $$

 It's not difficult to see that all the higher $T$-derivatives
$S_{TT}$, $S_{TTX}$, $S_{TTT}$, $\dots$, arising in the right-hand 
part can be replaces just by their main values defined by the 
Whitham system (\ref{ws1}) in this approximation.

\vspace{0.5cm}

 The work was partially supported by Russian Science Support 
Foundation.

\end{document}